\documentclass[5p]{elsarticle}
\usepackage{graphicx}
\usepackage{epsfig}

\usepackage{amssymb}
\usepackage{bm, bbm}  

\newcommand{\bone}{\mathbbm{1}}

\newcommand{\beq}{\begin{equation}}
\newcommand{\eeq}{\end{equation}}
\newcommand{\beqn}{\begin{eqnarray}}
\newcommand{\eeqn}{\end{eqnarray}}

\newcommand{\nn}{\nonumber}
\newcommand{\bq}{{\bf q}}

\newcommand{\bK}{{\bf K}}
\newcommand{\bA}{{\bf A}}

\newcommand{\bR}{{\bf R}}

\newcommand{\br}{{\bf r}}
\newcommand{\be}{{\bf e}}

\newcommand{\rhobar}{\bar{\rho}}
\newcommand{\Sbar}{\bar{S}}
\newcommand{\Ibar}{\bar{I}}

\newcommand{\ua}{\uparrow}
\newcommand{\da}{\downarrow}

\newcommand{\Fcal}{\mathcal{F}}
\newcommand{\Ocal}{\mathcal{O}}

\newcommand{\etab}{\mbox{\boldmath $\eta $}}

\begin{document}

\begin{frontmatter}

% Title, authors and addresses

% use the thanksref command within \title, \author or \address for footnotes;
% use the corauthref command within \author for corresponding author footnotes;
% use the ead command for the email address,
% and the form \ead[url] for the home page:
% \title{Title\thanksref{label1}}
% \thanks[label1]{}
% \author{Name\corauthref{cor1}\thanksref{label2}}
% \ead{email address}
% \ead[url]{home page}
% \thanks[label2]{}
% \corauth[cor1]{}
% \address{Address\thanksref{label3}}
% \thanks[label3]{}

\title{Theoretical expectations for a fractional quantum Hall effect in graphene}

% use optional labels to link authors explicitly to addresses:
% \author[label1,label2]{}
% \address[label1]{}
% \address[label2]{}

\author[mark,zlatko]{Z. Papi\'c}
\author[mark]{M. O. Goerbig}
\author[nicolas]{N. Regnault}

\address[mark]{Laboratoire de Physique des Solides, Univ. Paris-Sud,
CNRS, UMR 8502, F-91405 Orsay Cedex, France}
\address[zlatko]{Institute of Physics, P.O.Box 68, 11080 Belgrade, Serbia}
\address[nicolas]{Laboratoire Pierre Aigrain, Ecole Normale Sup\'erieure, 
CNRS, 24 rue Lhomond, F-75005 Paris, France}

\begin{abstract}

Due to its fourfold spin-valley degeneracy, graphene in a strong magnetic field may be viewed as a four-component
quantum Hall system. We investigate the consequences of this particular structure on a possible, yet unobserved,
fractional quantum Hall effect in graphene within a trial-wavefunction approach and exact-diagonalisation 
calculations. This trial-wavefunction approach generalises an 
original idea by Halperin to account for the SU(2) spin in semiconductor heterostructures with a relatively weak
Zeeman effect. Whereas the four-component structure
at a filling factor $\nu=1/3$ adds simply a SU(4)-ferromagnetic spinor ordering to the otherwise unaltered Laughlin 
state, the system favours a valley-unpolarised state at $\nu=2/5$ and a completely unpolarised state 
at $\nu=4/9$. Due to the similar behaviour of the interaction potential in the zero-energy graphene Landau level
and the first excited one, we expect these states to be present in both levels.

\end{abstract}

\begin{keyword}
% keywords here, in the form: keyword \sep keyword
Graphene \sep Strong Magnetic Fields \sep Fractional Quantum Hall Effect \sep Strongly Correlated Electrons

% PACS codes here, in the form: \PACS code \sep code
\PACS  73.43.-f \sep 71.10.-w \sep 81.05.Uw
\end{keyword}
\end{frontmatter}

% main text
\section{Introduction}
\label{intro}

Electrons in graphene may be viewed as a particular form of the two-dimensional
electron gas (2DEG), with the fundamental difference that, due to the particular
band structure, their low-energy properties are discribed in terms of a 
zero-mass Dirac equation rather than the usual effective-mass Schr\"odinger equation
\cite{antonioRev}. One of the most salient features of the 2DEG, when submitted to a
strong magnetic field, is the quantum Hall effect which occurs in an integer (IQHE)
as well as in a fractional (FQHE) form. The former is also manifest in graphene
\cite{novoselov,zhang}, and its obervation is a spectacular proof of relativistic
electrons (and holes) in graphene, due to an unusual quantisation of the
Hall conductivity, $\sigma_H=2(e^2/h)(2n+1)$, in terms of the integer $n$, as expected
on theoretical grounds \cite{ando,sharapov,PGCN}. 

Experimental evidence for the FQHE, which is due to electron-electron interactions in 
a partially filled Landau level (LL), is yet lacking in graphene. In the usual 2DEG in 
GaAs/AlGaAs heterostructures, the FQHE is, indeed, seen in samples with high mobilities yet 
unaccessed in graphene on a SiO$_2$ substrate ($\mu\sim 50\, 000\, {\rm cm^2/Vs}$
for typical samples). Higher mobilities ($\mu\sim 200\, 000\, {\rm cm^2/Vs}$)
have been achieved in current-annealed suspended graphene \cite{bolotin}, but unexpectedly
the IQHE happens to break down above 1T, probably due to extrinsic effects that 
are not related to the intrinsic electronic properties of these graphene samples
\cite{kim}. In spite of the missing FQHE, interaction physics is likely to be 
at the origin of additional plateaus in the Hall conductivity at LL filling
factors $\nu_G=\pm 1$ (and 0) \cite{zhang2}, where $\nu_G=n_{C}/n_B$ is the ratio between the 
carrier density ($n_C>0$ for electron and $n_C<0$ for hole transport) and that,
$n_B=B/(h/e)$ of the flux quanta threading the graphene sheet.

From a theoretical point of view, interactions in graphene LLs are expected to be 
relevant. Indeed, one needs to compare the typical energy for exchange interaction 
$V_C=e^2/\epsilon R_C\simeq 25 \sqrt{B[{\rm T}]}/\epsilon \sqrt{2n+1}$ meV, 
in terms of the dielectric constant $\epsilon$ and the 
cyclotron radius $R_C=l_B\sqrt{2n+1}$, with the magnetic length 
$l_B=\sqrt{\hbar/eB}=25/\sqrt{B[{\rm T}]}\, {\rm nm}$, to the LL separation
$\Delta_n=\sqrt{2}\hbar (v_F/l_B)(\sqrt{n+1}-\sqrt{n})$, where $v_F$ is the 
Fermi velocity. In spite of the decreasing LL separation in the large-$n$ limit,
the ratio between both energy scales remains constant and reproduces the
fine-structure constant of graphene, $\alpha_G=V_C/\Delta_n=e^2/\hbar v_F\epsilon
\simeq 2.2/\epsilon $. Notice that the Coulomb interaction respects the fourfold 
spin-valley degeneracy to lowest order in $a/l_B\simeq 0.005\sqrt{B[{\rm T}]}$,
where $a=0.14$ nm is the distance between nearest-neighbour carbon atoms in graphene.
This fourfold spin-valley symmetry is described in the framework of the SU($4$) 
group which covers the two copies of the SU(2) spin and the SU(2) valley isospin.
Lattice effects break this SU(4) symmetry at an energy scale 
$V_C (a/l_B)\simeq 0.1 B[{\rm T}]/\epsilon$ meV \cite{GMD,AF,Abanin,doretto}, which is
roughly on the same order of magnitude as the expected Zeeman effect in graphene
\cite{zhang2}. Other symmetry-breaking mechanisms have been proposed 
\cite{haldane,FL,herbut} but happen to be equally suppressed with respect to the leading
interaction energy scale $V_C$. An exception may be graphene on a graphite substrate,
where the natural lattice commensurability of the substrate and the sample may lead 
to a stronger coupling than for graphene on a SiO$_2$ substrate \cite{andrei}.
This yields a mass term in the Dirac Hamiltonian which lifts the valley degeneracy
of the zero-energy LL \cite{haldane}.

Based on these considerations, graphene in a strong magnetic field may thus be viewed 
as a four-component quantum Hall system, and we neglect SU(4)-symmetry breaking terms
in the remainder of the paper. An interesting theoretical expectation resulting
from this feature is the formation of a quantum Hall ferromagnet at $\nu=\pm 1$
\cite{nomura,GMD,AF,YDSMD} with SU(4)-skyrmion excitations, which may have peculiar
magnetic properties \cite{YDSMD,DGLM}. Also for the FQHE, the SU(4) spin-valley symmetry 
is expected to play a relevant role and has been considered within a composite-fermion
approach \cite{toke} as well as one based on SU(4) Halperin wavefunctions \cite{GR,DGRG}.

In this paper, we review how the four-component structure of graphene may have particular
signatures in a possible FQHE. In a first step, we discuss the structure of
the interaction model for electrons restricted to a single graphene LL. We concentrate
on the spin-valley SU(4) symmetric part of the interaction model, which constitutes the leading
energy scale, and discuss, based on the behaviour of the pseudopotentials, theoretical expectations
for the FQHE in graphene in the LLs $n=0$ and $1$. In the second part of this paper, we corroborate
these qualitative expectations with the help of the SU(4) wavefunction approach \cite{GR,DGRG} and
exact-diagonalisation calculations.

\section{Interaction model}

In the case of a partially filled LL, one may separate the ``low-energy'' degrees of freedom, which consist
of intra-LL excitations, from the ``high-energy'' inter-LL excitations. In the absence of disorder, all states 
within the partially filled LL have the same kinetic energy such that intra-LL excitations may be described by
considering only electron-electron interactions,
\beq\label{eq01}
H_{\lambda n}=\frac{1}{2}\sum_{\bq}v(q)\rho_{\lambda n}(-\bq)\rho_{\lambda n}(\bq),
\eeq
where $v(q)=2\pi e^2/\epsilon q$ is the 2D Fourier-transformed Coulomb interaction potential. The 
Fourier components $\rho_{\lambda n}(\bq)$ of the density operator are constructed solely from states within the
$n$-th LL in the band $\lambda$ ($\lambda=+$ for the conduction and $\lambda=-$ for the valence band). 
This construction is analogous to that used in the conventional 2DEG, but the density operators 
$\rho_n(\bq)$ are now built up from spinor states of the 2D Dirac equation,
\beq\label{eq02}
\psi_{\lambda n,m}^{\xi}=\frac{1}{\sqrt{2}}\left(\begin{array}{c} |n-1,m\rangle \\ \lambda |n,m\rangle \end{array}
\right)
\eeq
for $n\neq 0$ and 
\beq\label{eq03}
\psi_{n=0,m}^{\xi}=\left(\begin{array}{c} 0 \\  |n=0,m\rangle \end{array}
\right)
\eeq
for the zero-energy LL $n=0$, in terms of the harmonic oscillator states $|n,m\rangle$ and the guiding-centre 
quantum number $m=0$. Here, we have chosen the first component of the spinor to represent at the 
$K$ point ($\xi=+$) the amplitude on the $A$ sublattice and that on the $B$ sublattice at the $K'$ point
($\xi=-$). Notice that the valley and the sublattice indices happen to be the same in the zero-energy
LL $n=0$ and that, thus, a perturbation that breaks the inversion symmetry (the equivalence of the two
sublattices) automatically lifts the valley degeneracy \cite{haldane,FL,herbut}. In terms of the
spinor states (\ref{eq02}) and (\ref{eq03}), the density operator may be written
\beq\label{eq04}
\rho_{\lambda n}(\bq)=\sum_{\xi,m}\left(\psi_{\lambda n,m}^{\xi}\right)^{\dagger}e^{-i\bq\cdot\br}\psi_{\lambda n,m'}^{\xi}
c_{\lambda n,m;\xi}^{\dagger}c_{\lambda n,m';\xi}\ ,
\eeq
where $c_{\lambda n,m;\xi}^{(\dagger)}$ annihilates (creates) an electron in the state 
$\psi_{\lambda n,m}^{\xi}$. In Eq. (\ref{eq04}), we have neglected the contributions that are off-diagonal in
the valley index. Indeed, these contributions give rise to a rapidly oscillating phase $\exp(\pm i\bK\cdot\br)$ in
the matrix elements, where $\pm \bK=\pm (4\pi/3\sqrt{3}a)\be_x$ is the location of the $K$ and $K'$ points, respectively,
and yield terms in the Hamiltonian (\ref{eq01}), which break the valley-SU(2) symmetry of the interaction. They
are suppressed by a factor $a/l_B$ with respect to the leading interaction energy scale $e^2/\epsilon l_B$ \cite{GMD}, as mentioned
in the introduction. For the sake of simplicity and because of their smallness, we neglect these terms here.

Within the symmetric gauge, $\bA=(B/2)(-y,x,0)$, the position operator $\br$ in Eq. (\ref{eq04}) may be decomposed into 
the guiding-centre position $\bR$ and the cyclotron variable $\etab$. Whereas the latter only affects the quantum number
$n$, $\bR$ acts on $m$, and we may therefore rewrite the density operator (\ref{eq04}),
$\rho_{\lambda n}(\bq)=\Fcal_n(\bq)\rhobar(\bq)$,
as a product of the {\sl projected} density operator
\beq\label{eq06}
\rhobar(\bq)=\sum_{\xi;m,m'}\left\langle m\left|e^{-i\bq\cdot\bR}\right| m'\right\rangle c_{\lambda n,m;\xi}^{\dagger}
c_{\lambda n,m';\xi}
\eeq
and the {\sl graphene form factor}
\beqn\label{eq07}
\nn
\Fcal_n(q) &=& \frac{1}{2}\left(\left \langle n-1 \left|e^{-i\bq\cdot\etab}\right| n-1\right\rangle
+\left \langle n \left|e^{-i\bq\cdot\etab}\right| n\right\rangle\right)\\
\nn
&=& \frac{1}{2}\left[L_{n-1}\left(\frac{q^2l_B^2}{2}\right)+L_{n}\left(\frac{q^2l_B^2}{2}\right)
\right]e^{-q^2l_B^2/4}
\\
\eeqn
for $n\neq 0$, in terms of Laguerre polynomials, and 
\beq\label{eq08}
\Fcal_{n=0}(q)=\left \langle 0 \left|e^{-i\bq\cdot\etab}\right| 0\right\rangle=e^{-q^2l_B^2/4}
\eeq
for $n=0$. With the help of the projected density operators, the interaction Hamiltonian (\ref{eq01}) reads
\beq\label{eq09}
H_{\lambda n}=\frac{1}{2}\sum_{\bq}v_{n}^G(q)\rhobar(-\bq)\rhobar(\bq),
\eeq
where we have defined the effective interaction potential for graphene LLs,
\beq\label{eq10}
v_{n}^G(q)=\frac{2\pi e^2}{\epsilon q}\left[\Fcal_n(q)\right]^2.
\eeq

Notice that the structure of the Hamiltonian (\ref{eq09}) is that of electrons in a conventional 2DEG
restricted to a single LL if one notices that the projected density operators satisfy the magnetic
translation algebra \cite{GMP}
\beq\label{eq11}
[\rhobar(\bq),\rhobar(\bq')]=2i\sin\left(\frac{\bq'\wedge\bq l_B^2}{2}\right)
\rhobar(\bq+\bq'),
\eeq
where $\bq'\wedge\bq\equiv q_x'q_y-q_xq_y'$ is the 2D vector product. This is a remarkable result in view
of the different translation symmetries of the zero-field Hamiltonian; whereas the electrons in the 2DEG
are non-relativistic and therefore satisfy Galilean invariance, the relativistic electrons in graphene are
Lorentz-invariant. However, once submitted to a strong magnetic field and restricted to a single LL, the 
translation symmetry of the electrons is described by the magnetic translation group in {\sl both} cases.

\subsection{SU(4) symmetry}

The most salient difference between the conventional 2DEG and graphene arises from the larger internal symmetry of the latter, 
due to its valley degeneracy. This valley degeneracy may be accounted for by an SU(2) valley {\sl isospin} in addition to
the physical SU(2) spin, which we have omitted so far and the symmetry of which is respected by the interaction
Hamiltonian. Similarly to the projected charge density operator (\ref{eq06}), we may introduce spin and isospin 
density operators, $\Sbar^{\mu}(\bq)$ and $\Ibar^{\mu}(\bq)$, respectively, with the help of the 
tensor products \cite{DGLM}
\beqn\label{eq12}
\nn
\Sbar^{\mu}(\bq) &=& \left(S^{\mu}\otimes\bone\right)\otimes \rhobar(\bq),\\
\Ibar^{\mu}(\bq) &=& \left(\bone\otimes I^{\mu}\right)\otimes \rhobar(\bq).
\eeqn
Here, the operators $S^{\mu}$ and $I^{\mu}$ are (up to a factor $1/2$) Pauli matrices, which act on the spin
and valley isospin indices, respectively. The operators $(S^{\mu}\otimes\bone)$ and $(\bone\otimes I^{\mu})$
may also be viewed as the generators of the SU(2)$\times$SU(2) symmetry group, which is smaller than the
abovementioned SU(4) group. However, once combined in a tensor product with the projected density operators 
$\rhobar(\bq)$, the SU(2)$\times$SU(2)-extended magnetic translation 
group is no longer closed due to the non-commutativity of the 
Fourier components of the projected density operators. By commutating $[\Sbar^{\mu}(\bq),\Ibar^{\nu}(\bq)]$,
one obtains the remaining generators of the SU(4)-extended magnetic translation group \cite{DGLM}, which is,
thus, the relevant symmetry that describes the physical properties of electrons in graphene restricted to a single LL.

\subsection{Effective interaction potential and pseudopotentials}

Another difference, apart from the abovementioned larger internal symmetry, between the 2DEG and graphene 
in a strong magnetic field arises from the slightly different effective interaction potentials in the
$n$-th LL. The effective interaction for graphene is given by Eq. (\ref{eq10}) 
whereas that in the conventional 2DEG reads
\beq\label{eq13}
v_n^{2DEG}(q)=\frac{2\pi e^2}{\epsilon q}\left[L_n\left(\frac{q^2l_B^2}{2}\right)e^{-q^2l_B^2/4}
\right]^2.
\eeq
The difference between the two of them vanishes for $n=0$, as well as in the large-$n$ limit \cite{GMD}, but leads to quite 
important physical differences in the first excited LL ($n=1$) when comparing graphene to the 2DEG.

For the discussion of the FQHE, it is more appropriate to use Haldane's pseudopotential construction \cite{haldanePP},
which is an expansion of the effective interaction potential in the basis of two-particle states with a 
fixed relative angular momentum $\ell$. In the Laughlin state at filling factor $\nu=1/m$ \cite{laughlin},
\beq\label{eq14}
\phi_m^L(\{z_j,z_j^*\})=\prod_{k<l}(z_k-z_l)^m e^{-\sum_j|z_j|^2/4},
\eeq
in terms of the complex position $z_j=(x_j+iy_j)/l_B$ of the $j$-th particle, e.g., no particle pair has a 
relative angular momentum less than $\ell=m$. Therefore, all pseudopotentials $V_{\ell<m}$ are completely screened,
and the Laughlin state (\ref{eq14}) is the {\sl exact} $N$-particle ground state with zero energy of a 
model interaction potential with $V_{\ell<m}> 0$ and $V_{\ell\geq m}=0$ \cite{haldanePP}. Although this model
interaction potential is quite different from the pseudopotentials of the effective interaction potentials
(\ref{eq10}) and (\ref{eq13}),
it allows one to generate numerically the Laughlin state, which may be then compared to those obtained within
exact-diagonalisation calculations of the realistic interaction potential.

\begin{figure}
\centering
\includegraphics[width=6.5cm,angle=0]{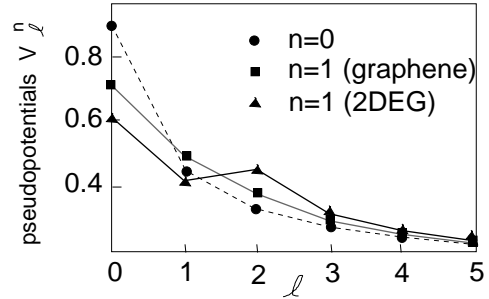}
\caption{\footnotesize{Pseudopotentials for graphene and the 2DEG in $n=0$ (graphene and 2DEG, circles), 
$n=1$ (graphene, squares), and $n=1$ (2DEG, triangles). 
The energy is measured in units of $e^2/\epsilon l_B$. The lines are a guide to the eye.
}}
\label{fig01}
\end{figure}

Notice that one may easily obtain the pseudopotentials of a given interaction 
potential $v_n(q)$ in Fourier space, such as (\ref{eq13}) or (\ref{eq10}), with the help of 
\beq\label{eq15}
V_{\ell}^n=\sum_{\bq}v_n(q) L_{\ell}(q^2l_B^2)e^{-q^2l_B^2/2}.
\eeq
The pseudopotentials for $n=0$ and $n=1$ in graphene and the 2DEG are shown in Fig. \ref{fig01}, which allows us
to make some qualitative statements about a potential FQHE in graphene as compared to that of the 2DEG. First, 
apart from the internal symmetry, the polarised FQHE states in the zero-energy LL are expected to be the same in 
graphene as in the 2DEG because there is no difference in the effective interaction potential. The only difference
stems from the larger internal symmetry in graphene, which affects the unpolarised FQHE states in $n=0$.
Second, the $n=1$ LL
in graphene is much more reminiscent of the $n=0$ LL than of the $n=1$ LL in the 2DEG. From an interaction point
of view, one would therefore expect that the quantum phases encountered in the $n=1$ graphene LL are merely a copy
of those in $n=0$. Furthermore, if one considers spin-polarised FQHE states, only pseudopotentials with {\sl odd}
angular momentum are relevant due to the fermionic nature of the electrons. It is apparent from Fig. \ref{fig01} that
odd-$\ell$ pseudopotentials are systematically larger in $n=1$ than in $n=0$. Therefore,
the overall energy scale of FQHE states in $n=1$ is slightly larger (by $\sim 10\%$) than in $n=0$, and one would expect, somewhat 
counterintuitively, that the $n=1$ FQHE states are more stable than those in $n=0$, for the same $B$-field value.
These qualitative predictions \cite{GMD} have been corroborated within exact-diagonalisation studies,
where only the valley isospin degree of freedom was considered and the physical spin was taken as completely polarised
\cite{AC,toke0}.

\section{Trial wavefunctions}

In order to account for the internal SU(4) symmetry in graphene LLs, two of us have proposed a trial-wavefunction
approach \cite{GR} based on an original idea by Halperin \cite{halperin} for the description of two-component 
FQHE states. These states,
\beq\label{eq16}
\psi_{m_1,...,m_K;n_{ij}}^{SU(4)}
=\phi_{m_1,...,m_4}^L \phi_{n_{ij}}^{inter} e^{-\sum_{j=1}^{K}
\sum_{k_j=1}^{N_j}|z_{k_j}^{(j)}|^2/4},
\eeq
consist of a product of four Laughlin wavefunctions (\ref{eq14}) (one per spin-valley component)
$$
\phi_{m_1,...,m_K}^L=\prod_{j=1}^4\prod_{k_j<l_j}^{N_j}\left(
z_{k_j}^{(j)}-z_{l_j}^{(j)}\right)^{m_j}
$$
and a term
$$
\phi_{n_{ij}}^{inter}=\prod_{i<j}^{4}\prod_{k_i}^{N_i}\prod_{k_j}^{N_j}
\left(z_{k_i}^{(i)}-z_{k_j}^{(j)}\right)^{n_{ij}}
$$
that takes into account correlations between the different components, labeled by the indices $i,j=1,...,4$.
Whereas the exponents $m_j$ must be odd integers to account for the fermionic statistics of the electrons,
the exponents $n_{ij}$, which describe inter-component correlations, may also be even. Notice further that 
not all wavefunctions are good candidates for a possible FQHE in graphene; it has indeed been shown, within
the plasma analogy \cite{laughlin}, that some wavefunctions correspond to a liquid in which the components
undergo a spontaneous phase separation \cite{DGRG}. 

The exponents $n_{ij}$ and $n_{jj}\equiv m_j$ define a symmetric $4\times 4$ matrix $M=(n_{ij})$, which
determines the component densities $\rho_j$ -- or else the component
filling factors $\nu_j=\rho_j/n_B$,\footnote{The filling factors used here are those that arise naturally in FQHE 
studies, i.e. they are defined with respect to the {\sl bottom} of the partially filled LL, in contrast to $\nu_G$ defined
with respect to the {\sl center} of $n=0$. In order to make the connection between the two filling factors, one needs to 
choose $\nu=\nu_1+ \nu_2+\nu_3 +\nu_4=\nu_G+2$. } 
\beq\label{eq17}
\left(\begin{array}{c} \nu_1 \\ \nu_2 \\ \nu_3 \\ \nu_4 \end{array} \right) = M^{-1}
\left(\begin{array}{c} 1 \\ 1 \\ 1 \\ 1 \end{array} \right).
\eeq
Eq. (\ref{eq17}) is only well-defined if the matrix $M$ is invertible. If $M$ is not 
invertible, some of the component filling factors, e.g. $\nu_1$ and $\nu_2$, remain unfixed, but not
necessarily the sum of the two ($\nu_1+\nu_2$). This is a particular feature of possible 
underlying ferromagnetic properties of the wavefunction \cite{GR}, as is discussed below for some
special cases.

In the following, we consider some particular subclasses of the trial wavefunctions (\ref{eq16}), which are 
natural candidates for a FQHE in graphene. Explicitely, we attribute the four spin-valley components as 
$1=(\ua,K)$, $2=(\ua,K')$, $3=(\da,K)$, and $4=(\da,K')$, where the first component denotes the spin 
orientation ($\ua$ or $\da$) and the second the valley ($K$ or $K'$). We investigate wavefunctions, where
all intracomponent exponents are identical $m_i=m$, i.e. we consider the same interaction potential for any
of the components, as it is the case in graphene. Furthermore, we consider $n_{13}=n_{24}\equiv n_{a}$ and
$n_{12}=n_{14}=n_{23}=n_{34}\equiv n_e$, which makes an explicit distinction between inter-component correlations in
the same valley ($n_a$) and those in different valleys ($n_e$). This distinction may occur somewhat 
arbitrary -- indeed, it does no longer treat the spin on the same footing as the valley isospin -- but it 
happens to be useful in some cases if one intends to describe states with intermediate polarisation, such
as for a moderate Zeeman field. The equivalence between spin and valley isospin is naturally 
restored for $n_e=n_a$. We use the notation $[m;n_e,n_a]$ to describe these subclasses of trial wavefunctions
(\ref{eq16}), the validity of which we check by exact diagonalisation of $N$ particles on a sphere \cite{haldanePP}.

\subsection{$[m;m,m]$ wavefunctions}

If all exponents are identical odd integers $m$, we obtain a completely antisymmetric orbital wavefunction, which
is nothing other than the Laughlin wavefunction (\ref{eq14}). In this case, the distinction between the
components vanishes, and the component filling factors are not fixed -- one may, without changing the orbital
wavefunction, fill only one particular component as well as another or distribute the particles over all
components. Only the total filling factor is fixed at $\nu=1/m$. The corresponding exponent matrix $M$ is
indeed not invertible (of rank 1), and the residual freedom of distributing the electrons over the four 
components may be viewed as the arbitrary orientation of a four-spinor in SU(4) space. The Laughlin wavefunction
in graphene is therefore associated with an SU(4) ferromagnetism, similar to that of the state
at $\nu_G=\pm 1$ \cite{nomura,GMD,AF,YDSMD,DGLM}, where a graphene quantum Hall effect has been observed at high
magnetic fields \cite{zhang2}.

As already mentioned above, the Laughlin wavefunction has the good property of screening all pseudopotentials
with angular momentum $\ell<m$ and has, for $m=3$, the usual large overlap with the Coulomb ground state \cite{toke}.

\subsection{$[m;m-1,m]$ wavefunctions}

A similarly good wavefunction is $[m;m-1,m]$, where the intervalley-component exponents are decreased by one. It also 
screens all pseudopotentials $V_{\ell<m}$ in any pair of electrons within the same valley, but an electron pair
in two different valleys is affected by the pseudopotential $V_{\ell=m-1}$. The filling factor, where this 
wavefunction may occur, is 
$$\nu=\frac{2}{2m-1},
$$
i.e. at slightly larger densities as the Laughlin wavefunction with the same $m$. 
The exponent matrix $M$ is still not invertible
but of rank 2, and indeed only the filling factors in the two valleys, $\nu_K=\nu_1+\nu_3$ and 
$\nu_{K'}=\nu_2+\nu_4$, respectively, are fixed, $\nu_K=\nu_{K'}=1/(2m-1)$. The wavefunction, thus, describes
a state with ferromagnetic spin ordering, but which is valley-isospin unpolarised. One may alternatively
view this $[m;m-1,m]$ wavefunction as a tensor product of an SU(2) Halperin $(m,m,m-1)$ isospin-singlet wavefunction 
\cite{halperin} and a completely symmetric (ferromagnetic) two-spinor that describes the physical spin.

\begin{center}\label{tab01}
\begin{table}
\begin{tabular}{c||c|c|c|c}
 number of particles $N$ & 4 & 6 & 8 & 10 \\
%\hline
% number of flux $2S$ & 7 & 12 & 17 & 22 \\
\hline
overlap $\Ocal$ in $n=0$ & 0.990 & 0.985 & 0.979 & 0.970 \\
\hline
overlap $\Ocal$ in $n=1$ & 0.965 & 0.882 & 0.896 & 0.876
\end{tabular}
\caption{\footnotesize{Overlap $\Ocal$ between the $(3,3,2)$ wavefunction and the state obtained from exact diagonalisation
of the effective interaction potential in $n=0$ and $n=1$. }}
\end{table}
\end{center}

We have checked within exact diagonalisation 
calculations that the $[3;2,3]$ wavefunction ($m=3$) describes indeed, to great accuracy, 
the ground state in graphene at $\nu=2/5$. It was shown by exact diagonalisation in Ref. \cite{toke} that, for
$N=4$ and 6 particles, the physical properties are indeed governed by an SU(2) symmetry, as suggested
by the $[m;m-1,m]$ wavefunction.
The overlap between this trial state and the one obtained by exact diagonalisation with implemented SU(2) symmetry
of the Coulomb interaction in $n=0$ and $1$ is shown in Tab. 1 for up to 10 particles. 
It is above 97\% for all studied system sizes in the 
zero-energy LL $n=0$, but slightly smaller ($\sim 88\%$) in $n=1$. We have used the planar pseudopotentials (\ref{eq15})
in the calculation of the $n=1$ LL and checked that the difference is less than 1\% in $n=0$ when compared to
using the more accurate spherical ones, even for the smallest system sizes $N=4$ and 6. 

It has been shown that
the ground state at $\nu=2/5$ in the conventional 2DEG is well described by an unpolarised $(3,3,2)$ SU(2)
Halperin wavefunction once the spin degree of freedom is taken into account \cite{2_5spinTheo}. This wavefunction
is identical to the composite-fermion wavefunction when including the SU(2) spin. The energy
difference between the polarised and the unpolarised $2/5$ states is, however, relatively small as compared
to the Zeeman effect at the corresponding magnetic fields, such that a polarised state is usually favoured.
Intriguing spin transitions have furthermore been observed experimentally at $\nu=2/5$ and hint to even more
complex physical properties of the 2/5 FQHE \cite{2_5spinExp}. Notice that the situation of the $[3;2,3]$ state
in graphene is remarkably different from that in the 2DEG: 
even in the presence of a strong Zeeman effect, only the ferromagnetically ordered physical spin 
is polarised, while the state remains a valley-isospin singlet. Whether such a valley-isospin singlet state
is indeed encountered in graphene depends sensitively on the valley-symmetry breaking terms; whereas a
possible easy-axis ferromagnetism, as has been proposed for the zero-energy LL $n=0$ \cite{AF}, may destroy
the $[3;2,3]$ state, it is favoured in the case of an easy-plane valley-isospin anisotropy, which may
occur in the $n=1$ graphene LL due to intervalley coupling terms of the order $V_C(a/l_B)$ \cite{GMD}.

We have furthermore studied the $(5,5,4)$ wavefunction ($m=5$) at $\nu=2/9$. Its overlap with the state obtained
by exact diagonalisation is lower than for the $(3,3,2)$ case (with $\Ocal=0.941$ for $N=4$ and 
$\Ocal=0.892$ for $N=6$), but remains relatively high.

\subsection{$[m;m-1,m-1]$ wavefunctions}

Another candidate is the $[m;m-1,m-1]$ wavefunction \cite{DGRG} which may describe FQHE states at 
$$\nu=\frac{4}{4m-3}.$$
The corresponding exponent matrix $M$ is now invertible, and the filling factor of each 
spin-valley component is $\nu=1/(4m-3)$, i.e. the state is an SU(4) singlet. As for the $[m;m,m]$ and
$[m;m-1,m]$ wavefunctions, all intracomponent correlations are such that the pseudopotentials 
$V_{\ell<m}$ are screened, but $V_{\ell=m-1}$ is relevant for all intercomponent interactions. 

As an example, we consider the $[3;2,2]$ wavefunction ($m=3$), which is a candidate for a possible graphene
FQHE at $\nu=4/9$. Our exact-diagonalisation calculations with implemented SU(4) symmetry, 
for $N=4$ and $8$ particles, indicate that 
this trial wavefunction describes indeed to great accuracy the ground state for the Coulomb interaction
potential in the $n=0$ LL, with an overlap of $\Ocal=0.999$ for $N=4$ and $\Ocal= 0.992$ for
$N=8$. In $n=1$, it is $\Ocal=0.944$ for $N=8$, for the case where one uses the
planar pseudopotentials (\ref{eq15}).
These results indicate that a possible $4/9$ FQHE state in graphene is, remarkably, of
a completely different nature than the composite-fermion state at $\nu=4/9$ in a one-component system,
such as the conventional 2DEG with complete spin polarisation. It is, nevertheless, an open issue to
what extent the SU(4) singlet state survives if one takes into account the Zeeman effect at high magnetic
fields, which favours a polarisation in the spin channel. A complementary composite-fermion 
calculation with SU(4) symmetry has revealed that, at $\nu=4/9$, states with intermediate SU(4) isospin
polarisation -- such as a valley-isospin singlet with full spin polarisation -- may exist, with a slightly 
higher energy than the composite-fermion SU(4) singlet \cite{toke}, which is indeed identical to 
the $[3;2,2]$ wavefunction. One may, therefore, expect a transition between
two $4/9$ states with different polarisation when the Zeeman energy outcasts the energy difference between the
two states. This is similar to the abovementioned $2/5$ state in a conventional 2DEG \cite{2_5spinTheo}.

\section{Conclusions}

In conclusion, we have investigated theoretically some particular features of the
FQHE in graphene as compared to the 2DEG. The electrons in graphene lose their relativistic
character associated with the Lorentz invariance once they are restricted to a single LL, in which
case the translations are governed by the magnetic translation group, as in the 2DEG case. 
The main difference between the 2DEG and graphene arises from the approximate SU(4) spin-valley symmetry,
which is respected in a wide energy range. Another difference arises from the spinor character of
the wavefunctions, which yields a different effective electron-electron interaction in graphene as
compared to the 2DEG. The graphene interaction potential in the 
first excited LL $n=1$ (in both the valence and the conduction band) is shown to be similar
to that in the central zero-energy LL $n=0$, yet with a slightly larger overall energy scale 
(roughly $10\%$ larger).

The FQHE at $\nu=1/3$ is described as a Laughlin state \cite{laughlin} with SU(4)-ferromagnetic
spin-valley ordering, similar to the state at $\nu=1$ \cite{nomura,GMD,AF,YDSMD,DGLM}. In 
contrast to this state, the system profits from its internal degrees of freedom by choosing a 
state with partial and full SU(4)-isospin depolarisation at $\nu=2/5$ and $\nu=4/9$,
respectively. The $[3;2,3]$ state at $\nu=2/5$ is a valley-isospin singlet, 
but its physical spin is ferromagnetically ordered and
may eventually be oriented by the Zeeman effect. The state at $\nu=4/9$ is described in terms
of a $[3;2,2]$ Halperin wavefunction, which is an SU(4) singlet with 
necessarily zero spin and valley isospin polarisation. A possible FQHE at $\nu=4/9$ in graphene may
therefore be sensitive to the Zeeman effect at high magnetic fields, and one may expect 
transitions between states with different polarisation, similar to the 2DEG at $\nu=2/5$ and
$2/3$ \cite{2_5spinExp}. 

\section*{Acknowledgments}

This work was funded by the Agence Nationale de la Recherche under Grant Nos. ANR-06-NANO-019-03
and ANR-JCJC-0003-01. ZP was supported by the European Commission, through a Marie Curie Foundation
contract MEST CT 2004-51-4307 and Center of Excellence Grant CX-CMCS, as well as by the Serbian 
Ministry of Science under Grant No. 141035. We acknowledge further material support from the 
Basque ``Les Bugnes'' Foundation.

% The Appendices part is started with the command \appendix;
% appendix sections are then done as normal sections
% \appendix

% \section{}
% \label{}

\end{document}